\documentclass{iopconfser}

\usepackage{amsmath}
\usepackage{graphicx}

\usepackage[bibstyle = ieee, citestyle = numeric-comp, url=false, isbn=false]{biblatex}
\begin{document}

\title{SHARP: A compact focusing system for medical applications using a diverging plasma lens}

\author{K N Sjobak$^1$, E Rød-Lindberg$^1$, A Ellingsen$^1$, P Drobniak$^1$, V F Rieker$^{1,2}$, F Reaz$^{3,4}$, C A Lindstrøm$^1$, and E Adli$^1$}

\affil{$^1$Department of Physics, University of Oslo, Oslo, Norway}
\affil{$^2$Center for Proton Therapy, Paul Scherrer Institute (PSI), Villigen, Switzerland}
\affil{$^3$Department of Clinical Medicine, Aarhus University, Aarhus, Denmark}
\affil{$^4$Danish Centre for Particle Therapy, Aarhus University Hospital, Aarhus, Denmark}
\email{k.n.sjobak@fys.uio.no}

\begin{abstract}
Cancer therapy for deep-seated tumors requires precise irradiation of a small target deep within the patient while minimizing radiation exposure to surrounding tissues.
This can be accomplished with a round beam sharply converging towards a single spot, requiring a large beam size in both planes at the exit of the focusing system.
Achieving this over a short distance using only quadrupole lenses is challenging; but by using a linear active plasma lens (APL) in defocusing mode, the beam can be quickly and non-destructively enlarged before focusing using quadrupoles.
The position of the irradiation spot can also be scanned in three dimensions by changing magnet settings.
The SHARP project will develop and test this concept.
Such a system can be used with very high energy electrons (hundreds of MeV), creating a Bragg-peak-like spot using novel accelerator technology.
This could lead to more compact radiotherapy facilities, not requiring a bulky infrastructure typically associated with proton radiotherapy machines.
If successful, SHARP will enable precision conformal radiotherapy, spatial fractionation, and potentially be useful for FLASH radiotherapy.
\end{abstract}

\section{Introduction}
The SHARP project~\cite{sjobak_particle_nodate} aims to design and demonstrate a novel and compact method for focusing charged particle beams to a small spot with a large opening angle, i.e.\ to a small $\beta^*$~\cite{rod-lindberg_compact_2024}, at some distance $L^*$ from the exit of the focus system as needed for patient positioning and target location inside the patient.
This method will achieve small transverse spot-sizes inside tissues, which are positioned at a precise depth defined by the focusing settings~\cite{reaz_sharp_2022}.
It is also natural to combine this with transverse positioning of the beam spot using steering magnets to achieve three-dimensional spot scanning.
The goal is to enable high-precision irradiation of cancer tumors, while avoiding depositing unwanted high doses to nearby healthy tissue.
Because the dose pattern is created through the geometrical convergence of the beam, the main limitation for the minimum transverse spot size is multiple Coulomb scattering, which disturbs this convergence.
The technique can be applied to both electrons and protons; for protons it can also be combined with energy modulation to combine the Bragg peak and focal spot, further enhancing the peak dose~\cite{reaz_sharp_2022}.
For electrons, it makes possible the creation of Bragg-peak-like dose geometry~\cite{kokurewicz_focused_2019,kokurewicz_experimental_2021,whitmore_focused_2021,whitmore_cern-based_2024}, potentially enabling the use of smaller and simpler accelerators for precision particle therapy than with protons, while also benefiting from the rapid development of novel accelerators using plasma and high-gradient radiofrequency accelerating structures~\cite{fischer_very_2024,angoletta_deft_2025,svendsen_focused_2021,eaac_seimetz_laser,eaac_giaccaglia_temporalBio,eaac_svensson_compact}.

A novel component in our proposed implementation is the use of an Active Plasma Lens (APL), which is a magnetic device that focuses or defocuses the beam~\cite{van_tilborg_active_2015,panofsky_focusing_1950}.
Unlike a quadrupole, the effect on the horizontal and vertical plane of the beam dynamics are equal, acting on the beam particles with a magnetic Lorentz force which is linearly increasing with radius, either focusing or defocusing the beam depending on the direction of the electrical current which is driven through the device.
This electrical current creates a current density $j_z$ in the beam aperture $r<R$, so that in this region, Ampère's law can be expressed as
\begin{equation}
    \nabla \times \vec B = \mu_0 \vec j \Rightarrow \mu_0 j_z = \frac{\partial B_y}{\partial x} - \frac{\partial B_x}{\partial y} \;,
    \label{eq:Faraday}
\end{equation}
where $\{x,y,z\}$ are coordinates in a right-handed coordinate system and $r=\sqrt{x^2+y^2}$.
For convenience, we define the magnetic gradient in each plane as
\begin{equation}
    g_x \equiv \frac{\partial B_y}{\partial x} \,,\; g_y \equiv - \frac{\partial B_x}{\partial y}\;.
    \label{eq:gradients-def}
\end{equation}
Furthermore, we can define the lens current as
\begin{equation}
    I_\mathrm{APL} \equiv 2\pi \int_0^R j_z(r) \,r\,\mathrm{d}r \;,
    \label{eq:Iapl-j}
\end{equation}
and using the integral form of Ampère's law, when $j_z(r)$ is constant for $0 < r < R$, we have that the circumferential magnetic field generated by the internal current $j_z$ is
\begin{equation}
    B_\theta(r) = \frac{\mu_0 r I_\mathrm{APL}}{2\pi R^2}\;,
\end{equation}
where
\begin{equation}
    \vec B (r,\theta) = B_\theta (r) \hat \theta = B_\theta (t) \left(-\sin(\theta)\hat x + \cos(\theta)\hat y\right) = B_x(x,y) \hat x + B_y(x,y) \hat y\;.
\end{equation}
Thus, the components of the magnetic field are:
\begin{align}
    B_x = -\sin\theta  \frac{\mu_0 r I_\mathrm{APL}}{2\pi R^2} 
        &= - y \frac{\mu_0 I_\mathrm{APL}}{2\pi R^2} = - y g_y \label{eq:Bx}\\[8pt]
    B_y =  \cos\theta  \frac{\mu_0 r I_\mathrm{APL}}{2\pi R^2}
        &= x \frac{\mu_0 I_\mathrm{APL}}{2\pi R^2} = x g_x \label{eq:By}
\end{align}
Here, $\tan\theta = y/x$, and $\{\hat \theta, \hat r, \hat x, \hat y, \hat z \}$ are unit vectors along their respective axes.
The plasma lens gradients were found by applying Equation~\eqref{eq:gradients-def} to $B_x$ and $B_y$ in Equations~\eqref{eq:Bx}--\eqref{eq:By} and then substituting in the result, giving
\begin{equation}
    g \equiv g_x = g_y = \frac{\mu_0 I_\mathrm{APL}}{2\pi R^2}\,.
\end{equation}
When $\vec j = j_z \hat z = \hat z\, I_\mathrm{APL}/\left(\pi R^2\right)$ is constant within the aperture and the velocity of the beam particles with charge $q$ is $\vec v = \hat z \, v_z$, the Lorentz force on the beam particles from the magnetic field becomes
\begin{equation}
    \vec F = q \vec v \times \vec B = - \hat r \, q v_z B_\theta = - \hat r \, q v_z \, g r = - \hat r \, q v_z \, \frac{\mu_0 I_\mathrm{APL}}{2\pi R^2} r\,.
\end{equation}
Thus, for a positively charged beam traveling in the $+\hat z$ direction, a positive electric current density $j_z$ results in a focusing force in the $-\hat r$ direction, and a defocusing force on a negatively charged beam.
Note that since Ampère's law is linear, field components from other currents may superimpose with the field generated by the assumed constant current density $j_z$, and contribute to the total magnetic field.
An example of this occurs inside the beam aperture of conventional accelerator quadrupole magnets, where $\vec j = 0$, so that $g_x + g_y = 0$.
For quadrupoles and plasma lenses, we often express the strength in terms of normalized focusing strength $k$ as used in Hill's equation, rather than directly in terms of magnetic gradient.
This is defined as $k=\pm g q / p_0$, where $p_0$ is the reference momentum of the beam.
It should be noted that both positive and negative conventions for the sign of $k$ are used by various authors, e.g.~\cite{wille_physics_2005,appleby_science_2020}.

\begin{figure}[tbp]
    \centering
    \includegraphics[width=0.48\linewidth]{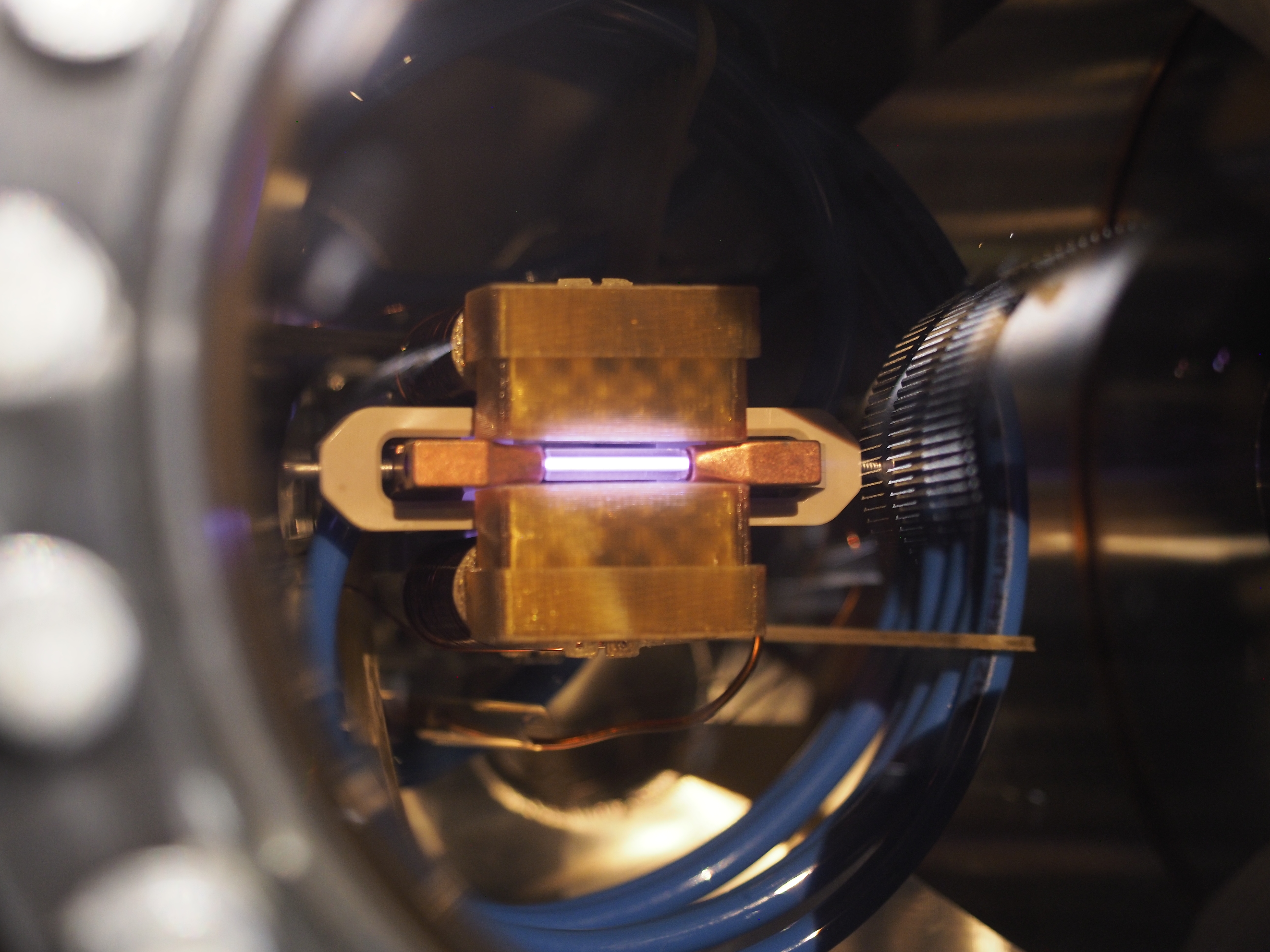}\hfill
    \includegraphics{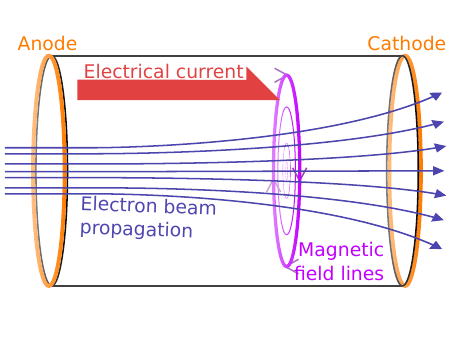} 
    \caption{Left: Photograph of the Active Plasma Lens (APL) at CLEAR, while discharging in with argon. Right: Principle of operation for an APL, showing the direction of the electrical current in the plasma driven by the APL's power source, and the resulting magnetic field and its effect on a negatively charged beam.}
    \label{fig:APLpic}
\end{figure}

We are investigating the properties of APLs at the Plasma Lens Experiment test-stand~\cite{lindstrom_overview_2018} at the CERN Linear Electron Accelerator for Research (CLEAR) facility~\cite{gamba_clear_2018}, as part of the CLEAR Plasma Lens Experiment collaboration.
We have previously demonstrated that APLs can have a linear focusing force while preserving beam emittance~\cite{lindstrom_emittance_2018}.
Additionally we have shown that they can reach very high focusing gradients of at least 3.6~kT/m~\cite{sjobak_strong_2021}, while maintaining linearity and allowing for tunable strength.
The APL at CLEAR, illustrated in Figure~\ref{fig:APLpic}, is a plasma discharge device driven by a strong current source, such as a Compact Marx Bank~\cite{dyson_compact_2016}.
The strength of the plasma lens is tuned simply by changing the current in the lens $I_\mathrm{APL}$ at the time of beam arrival.

\begin{figure}[p]
    \centering
    \includegraphics{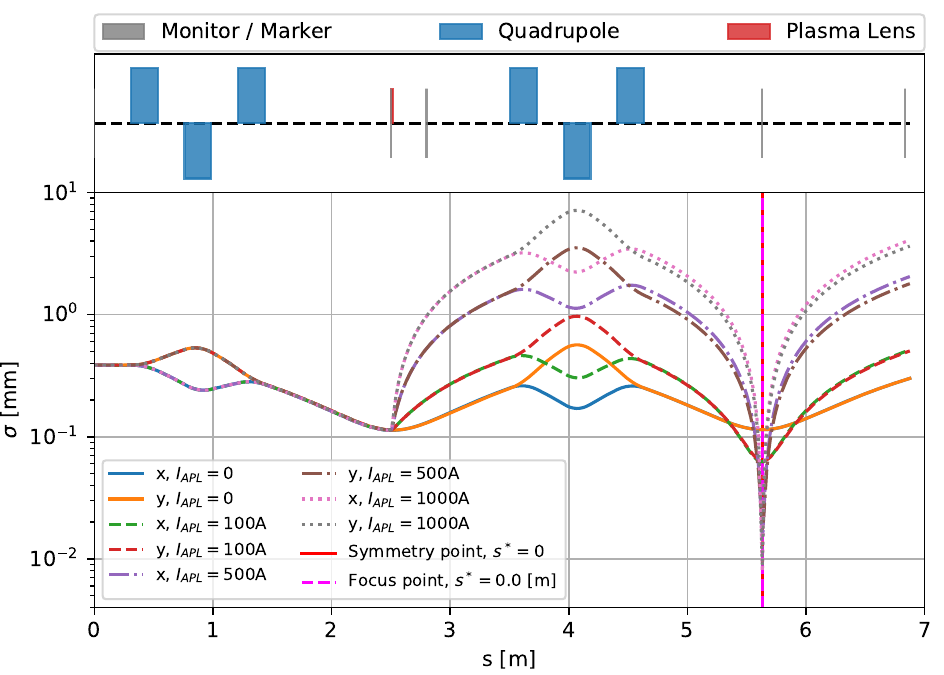}
    \vspace{-11pt} 
    \includegraphics{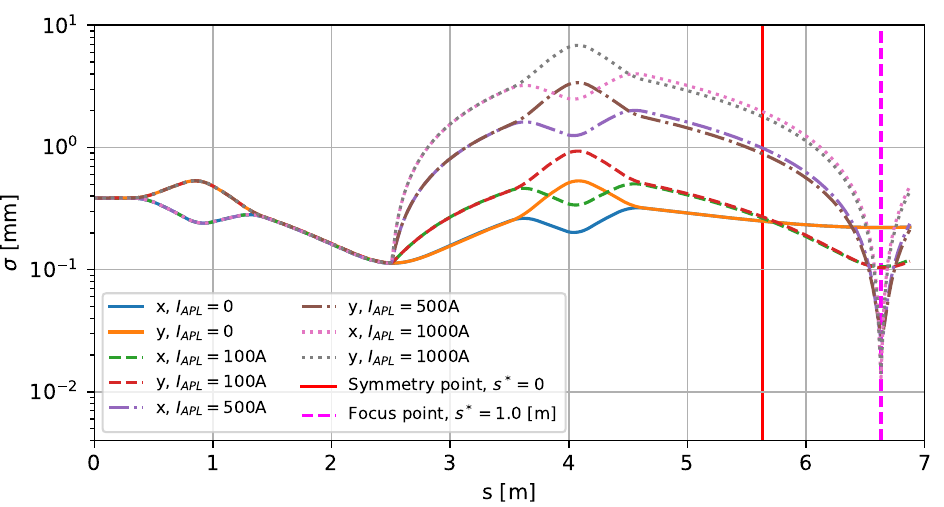}
    \caption{
    Simulated beam size ($\sigma$) in the horizontal ($x$) and vertical ($y$) plane as a function of distance along the beam line ($s$), with SHARP focusing optics.
    Log scale is used for the vertical axis to clearly see the depth of the focus point. 
    Top plot: Focus at symmetry point of the APL relative to the center of the final focusing triplet, at $s^*=0$, and beamline layout above.
    Bottom plot: Focus 1~m behind the symmetry point, at  $s^*=1$~m. Both plots show four cases of the APL current $I_\text{APL}$: Off ($I_\text{APL} = 0$), low (100~[A]), moderate (500~[A]), and high (1000~[A]).
    }
    \label{fig:APLsigma}
\end{figure}

The operating principle of the proposed SHARP hybrid focus system is to first expand the beam in an APL before focusing it with a quadrupole triplet.
If needed, additional quadrupole magnets can be used to match the beam from the accelerator into the APL.
A simulation illustrating this principle, which is described in more detail in the next section, is shown in Figure~\ref{fig:APLsigma}.
The use of an APL in SHARP enables a compact system, as it can quickly expand the beam --- which is a needed step in the focusing process --- without scattering it.
Compared to a pure quadrupole setup, this design allows for a system in which both the focal point position and convergence angle can be tuned, naturally producing a round beam after exiting the final focus quadrupoles.

The introduction of the defocusing linear APL in the lattice reduces the Twiss $\beta$-function~\cite{wille_physics_2005} $\beta^*$ at the focal point, since the evolution of the $\beta$-function in the field-free area around the focal point is given by
\begin{equation}
    \beta(s) = \frac{s^2}{\beta^*} + \beta^*\;,
    \label{eq:beta-s}
\end{equation}
where $s$ is the distance from the focal point.
The beta function is related to the beam size in a single plane as $\sigma=\sqrt{\beta\epsilon_g}$, where $\epsilon_g$ is the geometric beam emittance, which remains constant in the absence of imperfections and acceleration.
The beta function at the focal point also controls the convergence angle of the beam towards that location, since $\sigma' = \sqrt{\epsilon_g/\beta^*}$.
From this we can deduce that for $\beta^* \ll s$, $\beta(s)$ must become large.
This is the case at the exit of the final focus quadrupoles, i.e.\ at $s=L^*$.
Note that to have a small focus point, this inequality must be true, since some distance $L^*$ between the equipment and the patient is needed.
Furthermore, inside quadrupoles and quadrupole assemblies of moderate length, the change in $\beta$ is limited; their main effect is changing its derivative $\alpha \equiv - 0.5 \frac{\mathrm{d}\beta}{\mathrm{d}s}$.
Thus, $\beta^*$ is to a large extent driven by the $\beta$ function at the entrance of the final focusing quadrupoles.
The role of the APL in the SHARP hybrid focus system is to create this large $\beta$-function at the quadrupole entrance, enabling a small focal spot with sharply converging beam, as described in Section~\ref{sec:SHARP-focus}.

A similar technique is described by Lan et al.~\cite{lan_experimental_2023}, using a single plasma lens with two sections, the first half of the lens defocusing and the second half focusing.
However, this design sacrifices flexibility and performance for reduction of complexity, as only a single parameter --- the APL current --- controls both the distance to the focusing spot $L^*$ and the beam's opening angle out of the defocusing section.
Additionally, it reduces performance by eliminating the drift between the sections where the beam can expand.
Also, the focusing APL aperture limits $\beta(L^*)$, therefore restricting the maximum effective $L^*$ if a small $\beta^*$ is to be reached.

Note that if $s \ne 0$ and $\beta(s)$ is specified, Equation~\eqref{eq:beta-s} has two solutions~\cite{fardous_reaz_advanced_2021} for $\beta^*$: One corresponding to a sharp focus where $\beta^* \ll s$, as discussed above, and another referred to as the ``high-$\beta$''~\cite{burkhardt_high-beta_2019} solution where $\beta^* \approx \beta(s)$, and the particles travel approximately parallel to each other.
This is discussed in Section~\ref{sec:SHARP-bigbeam}, and makes it possible to have a large beam with an almost constant size after exiting the final focus quadrupoles.
This can be created using the same magnetic beamline elements in the same positions as for when sharply focusing the beam, only changing their field strengths.
This type of beam is potentially very interesting for FLASH radiotherapy~\cite{vozenin_towards_2022} --- a modality that aims to spare healthy tissue through ultra-fast beam delivery --- as it would allow to simultaneously irradiate a large volume within a single beam pulse.
The alternative methods for beam enlargement, such as pencil beam scanning, prolong the irradiation time~\cite{folkerts_framework_2020}, while scattering~\cite{robertson_demonstration_2026} reduces the beam intensity and produces stray radiation.

\section{Simulation of SHARP hybrid focus system}
\label{sec:SHARP-focus}
Working backwards starting from the focus point inside the patient, the final focusing system must have an appropriately sized aperture that is large enough to accommodate a strongly focused beam towards the target point with a large opening angle, while also ensuring sufficient separation between the beamline equipment and the patient.
A quadrupole triplet is the most suitable option to accommodate this requirement, as a very large-aperture APL would be technically difficult in terms of peak current and gas flow (proportional to aperture area), as well as skin depth and Z-pinch effects potentially making it difficult to achieve linear fields~\cite{christiansen_studies_1984,bennett_magnetically_1934,rosenbluth_infinite_1955,sjobak_strong_2021}.
Quadrupole multiplets are also a well known solution for beam focusing.

A system for beam expansion is used to produce the beam that is received by the final focusing system.
The SHARP concept uses an APL for this, with the current polarity set to the defocusing direction, i.e.\ $I_\mathrm{APL} < 0$ in the sense of Equation~\eqref{eq:Iapl-j}; in the following discussion the sign is dropped.
This configuration enables direct defocusing in both planes simultaneously, providing a tunable and compact beam expansion system, which is then paired with the final focusing system.
Note that there should be a short field-free drift section between the APL exit and the quadrupole entrance, in order to let the beam expand.

A matching triplet is used to accept the incoming beam from the accelerator and transform it to a size appropriate for the APL.
Here, the initial beam is assumed to be a 200~MeV electron beam, with a normalized emittance of 10~mm~mrad.
The matching triplet strengths are set to create a round beam waist in the center of the APL when the APL is off, with $\beta = 0.5$~m and $\sigma = 0.11$~mm.

At a distance of 31~cm before the entry to the matching triplet, the $\beta$ function in both planes is assumed to be 5.8163~m, and $\alpha=0$.
These are representative parameters for the CLEAR accelerator~\cite{rod-lindberg_compact_2024}.
The quadrupole magnets in both triplets are assumed to be 22.6~cm long, and have a spacing of 22.4~cm.
The APL has a length of 20~mm and a diameter of 1~mm, and its center is 1.172~m after the exit of the matching triplet, and 1.174~m before the entry of the final focus triplet.
To simplify the description of the focal point, a coordinate system $s^*$ is sometimes used, defined such that $s^* \equiv 0$ at the ``symmetry point''. This point lies at a distance from the exit of the focusing triplet equal to the distance between the APL center and the focusing triplet entrance.

The output from a linear optics simulation using ImpactX~\cite{huebl_next_2022} version~25.12 for this system is displayed in Figure~\ref{fig:APLsigma}, showing how the beam envelope evolves throughout the lattice without energy spread and other imperfections.
A combination of four APL currents giving different defocusing strengths are shown, for two different focal point positions.
From this graph, it is clear that engaging the APL, even at moderate currents, makes the beam envelope cones expand quickly around the focal point.
These cones represent the volumes traversed by the beam towards and away from a beam waist.
This increase in cone size drastically enhances the focusing performance in the selected focal spot.
We also see that it makes much longer focal lengths achievable with good performance.
Additionally, the angle of convergence of the beam envelope cone is much larger, causing the beam to spread out rapidly on either side of the focal point, which reduces the dose before and after the focal point.

\begin{figure}[tbp]
    \centering
    \includegraphics{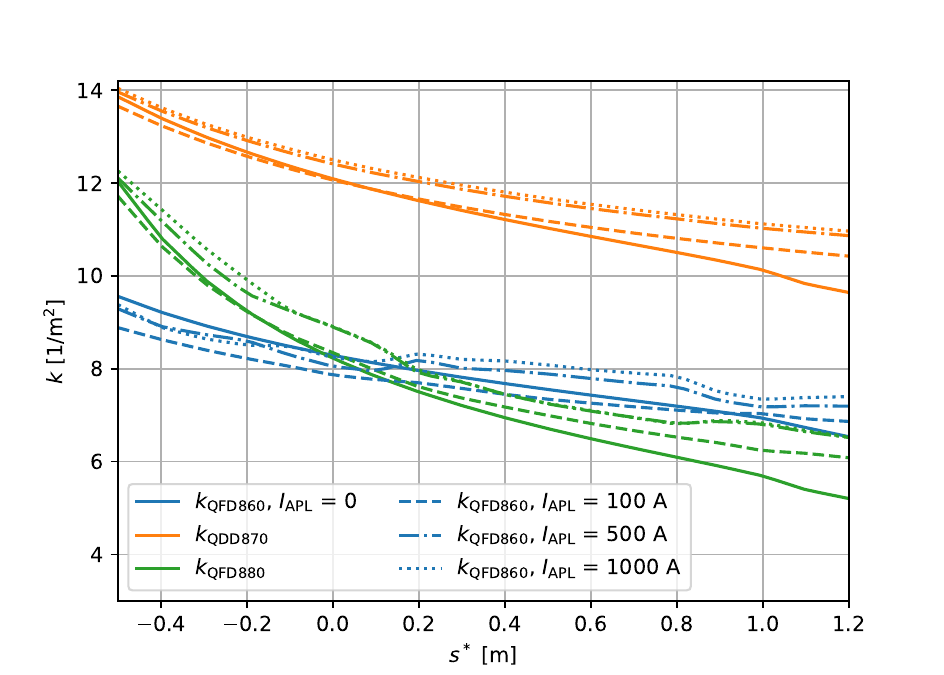}
    \caption{Quadrupole strengths in the final focus triplet as a function of different focal point position $s^*$, for different APL currents, optimized by minimizing Equation~\eqref{eq:r-quads}. The first column of the legend shows how the different quadrupole magnets are represented by different colors, and the right column shows how different APL currents correspond to different line dash patterns.}
    \label{fig:k-quads}
\end{figure}

The strengths of the final focus triplet quadrupoles are optimized by minimizing the residual function $r(k_\mathrm{QFD860},k_\mathrm{QDD870},k_\mathrm{QFD880})$ given by~\cite{rod-lindberg_compact_2024}
\begin{equation}
    r = 1000*(\alpha_x^2 + \alpha_y^2) + (\beta_x-\beta_y)^2 \,,
    \label{eq:r-quads}
\end{equation}
where $\alpha_x$, $\alpha_y$, $\beta_x$, and $\beta_y$ are the Twiss parameters in the desired focal spot.
The variables $k_\mathrm{QFD860}$, $k_\mathrm{QDD870}$, and $k_\mathrm{QFD880}$ represent the strengths of the three final focus quadrupoles seen in the beamline layout drawing in the top of Figure~\ref{fig:APLsigma}.
If the beam has a waist at this focal spot and is round, then $r=0$.
The factor 1000 was inserted to prioritize having a waist in both planes over a round beam, and also taking into account that the numerical values of $\alpha$ and $\beta$~[m] have different units.
The resulting quadrupole strengths as a function of $s^*$, for four different values of $I_\mathrm{APL}$, are shown in Figure~\ref{fig:k-quads}.
For the optimization, the Nelder-Mead miminization algorithm in SciPy version~1.16.3 was used~\cite{virtanen_scipy_2020}, with a tolerance setting of $10^{-5}$.

\begin{figure}[tbp]
    \centering
    \includegraphics{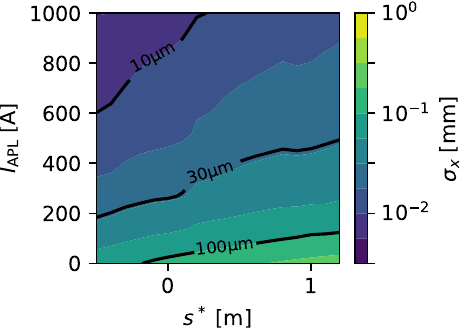}
    \hfill
    \includegraphics{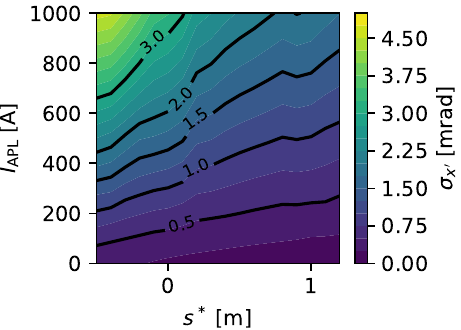}
    \caption{Beam size $\sigma$ and opening angle $\sigma'$ in the focal spot, as a function of focus position $s^*$ and APL current $I_\mathrm{APL}$. Data for the horizontal ($x$) plane is shown; the vertical plane is very similar. This simulation assumes no scattering, linear optics, and Gaussian beams.}
    \label{fig:sigmasMap}
\end{figure}

The beam size and opening angle at the focal point is shown in Figure~\ref{fig:sigmasMap}.
These plots show that increasing the APL current reduces the beam size while increasing the opening angle for all focal distances.
Since the beam is round, only the horizontal plane is shown as the vertical plane is virtually identical.
Note that the color scale for the beam size is logarithmic, in order to show the gradual approach towards an infinitely small beam for high $I_\mathrm{APL}$ and low $s^*$.
In contrast, the color scale for the opening angle is linear, showing the gradual increase at high $I_\mathrm{APL}$ and low $s^*$.

\begin{figure}[tbp]
    \centering
    \includegraphics{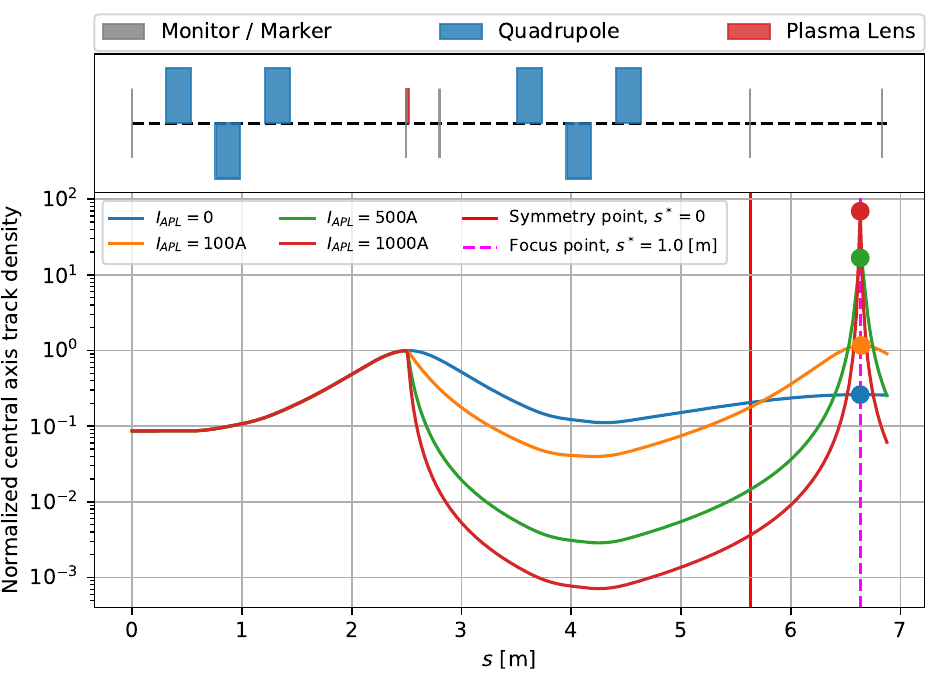}
    \caption{Particle fluence as given in Equation~\eqref{eq:fluence}, normalized to 1.0 at the peak inside the APL, for four different APL currents. The focal point is located at $s^* = 1.0$~m, and the peaks of the fluence are indicated with circles of the same color.}
    \label{fig:trackDensity}
\end{figure}

The central axis particle fluence can be estimated as
\begin{equation}
    \Phi = \frac{N}{\pi\sigma_x\sigma_y}\,,
    \label{eq:fluence}
\end{equation}
which is the number of particles $N$ divided by the area of a $1 \sigma$ ellipse of the beam.
This fluence is shown as a function of $s$ for different APL currents in Figure~\ref{fig:trackDensity}.
As expected, the particle fluence is strongly affected by the APL, with a pronounced spike appearing in the focal point when it is active.
In contrast, without the APL there is only a minimal peak, unless $s^*$ is very short.
This spike is also very sharp, falling off rapidly behind and in front of the selected focal point.
In the plot, we have normalized the curves so that the peak fluence in the APL is 1.0.

\begin{figure}[tb]
    \centering
    \includegraphics{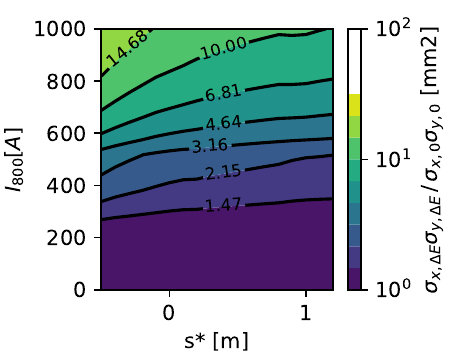}
    \hfill
    \includegraphics{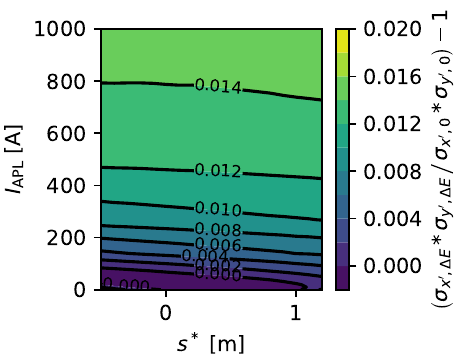}
    \caption{Relative beam area $\pi\sigma_x\sigma_y$ (left) and opening angle product $\sigma_{x'}\sigma_{y'}$ (right), comparing tracking simulations where $\sigma_{p_t} = 0.01$ with envelope simulations without energy spread.}
    \label{fig:sigmasRelMap}
\end{figure}

\begin{figure}[tb]
    \centering
    \includegraphics{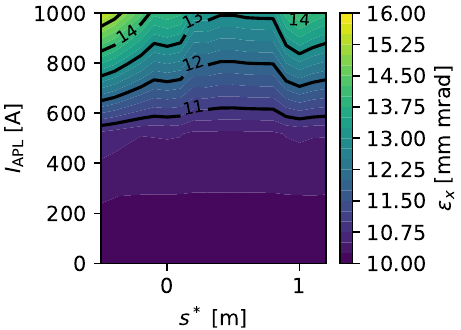}
    \hfill
    \includegraphics{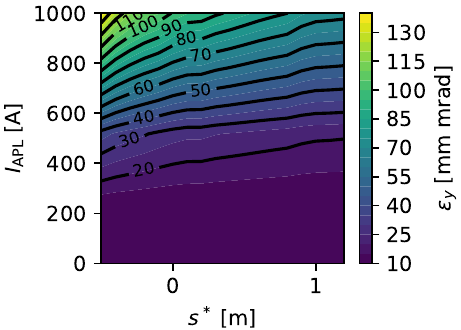}
    \caption{Normalized emittance at the focus point for the horizontal (left) and vertical (right) plane, from tracking simulations where $\sigma_{p_t} = 0.01$ and the initial normalized emittance is $10~[\mathrm{mm\,mrad}]$.}
    \label{fig:emittanceMap}
\end{figure}

To investigate the effect of energy spread, tracking simulations with a Gaussian-distributed relative momentum offset with standard deviation $\sigma_{p_t} = 0.01$, i.e.\ 1\%, were performed for each point in the parameter space $\left(s^*,I_{APL}\right)$, using the previously optimized quadrupole strengths.
Here, $p_t = -\Delta(\gamma)/(\beta_0\gamma_0)$, where $\beta_0$ and $\gamma_0$ are the relativistic factors associated with the reference velocity, and $\Delta(\gamma)$ the deviation from the reference relativistic factor $\gamma$~\cite{huebl_next_2022}.
The results from these simulations are shown in Figure~\ref{fig:sigmasRelMap}, which combines the impact in both planes, and in Figure~\ref{fig:emittanceMap} which shows the emittance increase in each plane.
From the plots, it is clear that the energy spread mainly affects the beam size at the focal point, while the angular opening remains approximately constant.
When examining the planes separately, we see that the impact on the beam size is nearly ten times larger in the vertical plane than in the horizontal plane.
This observation is also reflected in the emittance (Figure~\ref{fig:emittanceMap}), where the effect is much larger in the vertical plane than in the horizontal.
Additionally, the normalized emittance is increasing slowly for $I_\mathrm{APL}$ in the range 0 to 300~A regardless of $s^*$.
The improvement of both $\sigma$ and $\sigma'$ within this range of $I_\mathrm{APL}$ is substantial, as shown in Figure~\ref{fig:sigmasMap}.

\section{Direct optical beam enlargement}
\label{sec:SHARP-bigbeam}
Another potential application of the combination of a diverging APL and final focus quadrupoles is beam enlargement via purely optical means, i.e.\ without scattering, which is the currently used technique at CLEAR for creating large-area and also uniform beams~\cite{robertson_demonstration_2026}.
Such beams are essential for FLASH radiotherapy, as it is needed to irradiate the entire target very quickly.
Large beams are also very useful for radiobiology and other radiation effects research using small targets, as a larger field reduces dose uncertainty from positioning in non-uniform (Gaussian) transverse beam distributions.

For such applications, we can adjust the strengths of the final focus quadrupole magnets to create the ``high-$\beta$'' solution of Equation~\eqref{eq:beta-s} mentioned in the introduction.
These solutions are also found by minimizing Equation~\ref{eq:r-quads}, however a different set of initial parameters were used for the quadrupole strengths, found by hand-tuning them to approximately the wanted solution before running the minimization algorithm.
This approach results in a different local minimum for $r\left(k_\mathrm{QFD860}, k_\mathrm{QDD870}, k_\mathrm{QFD880} \right)$.

\begin{figure}[tbp]
    \centering
    \includegraphics{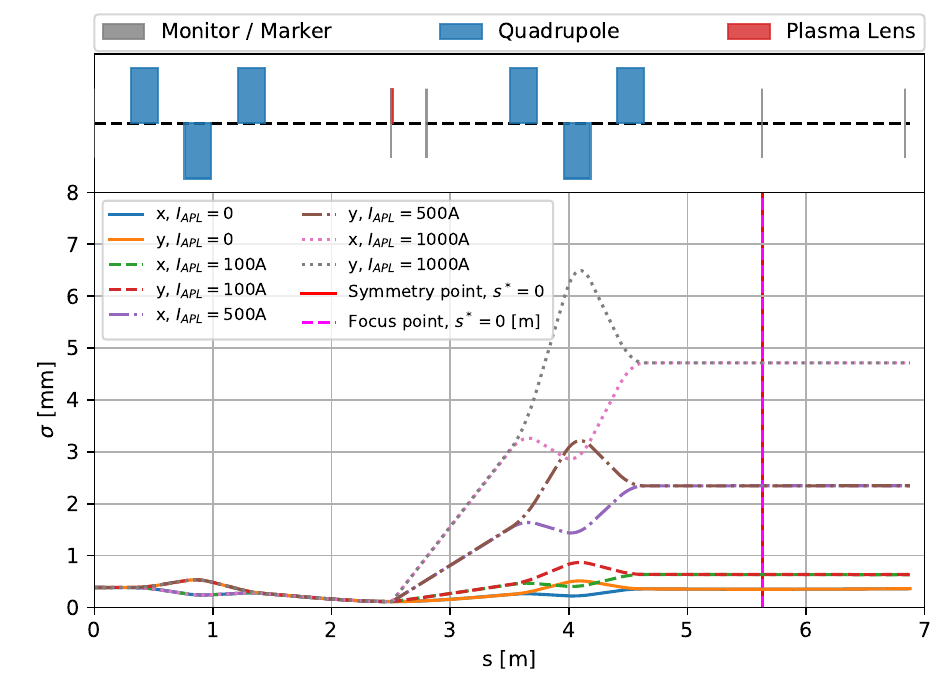}
    \caption{Beam size ($\sigma$) in the horizontal and vertical plane as a function of distance ($s$) along the beam line, for the direct enlargement case.}
    \label{fig:APLsigma2}
\end{figure}

\begin{figure}[tbp]
    \centering
    \includegraphics{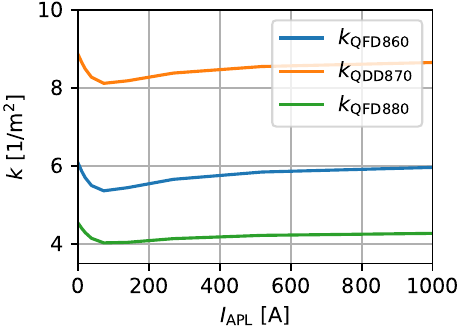}
    \hfill
    \includegraphics{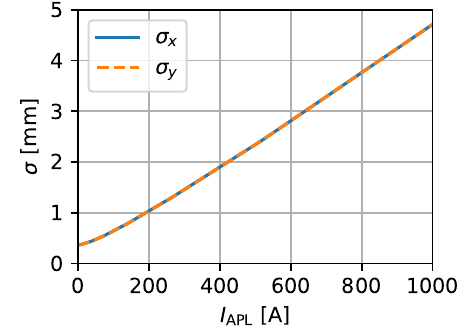}
    \caption{Parameters for a large parallel ``high-$\beta$'' beam. (Left) Final focus triplet strengths to ``catch'' and make the beam parallel and round, as a function of plasma lens current. (Right) Beam size at $s^*=0$, for both planes, as a function of plasma lens current.}
    \label{fig:bigParams}
\end{figure}

The resulting beam size evolution along the beam line for four different APL currents $I_\mathrm{APL}$ is shown in Figure~\ref{fig:APLsigma2}, while the magnet strengths used to create this are shown in Figure~\ref{fig:bigParams}~(left).
The resulting beam size at $s^*=0$, which is identical in both planes and essentially constant after leaving the final quadrupole magnet, is shown in Figure~\ref{fig:bigParams} (right).


\section{Discussion}

From the hybrid focus system simulation results, it is evident that the system operates effectively, creating a small, round, and movable focus point with sharp convergence angles.
We also see that the focusing system performs well also in the presence of beam energy spread.
While the parameters presented in this paper are meant to be illustrative and loosely based on the CLEAR accelerator, the overall conclusions should be broadly valid.

The focal point position $s^*$ is adjusted by changing the quadrupole strengths, as shown in Figure~\ref{fig:k-quads}.
It is evident that the quadrupole strengths depend relatively little on the APL current. 
When focused at the symmetry point $s^* = 0$, the first and last quadrupoles have equal strength, as expected from symmetry.

The much larger effect that energy spread has on the vertical emittance, as shown in Figure~\ref{fig:emittanceMap}, can be understood from Figure~\ref{fig:APLsigma}.
These plots show that the ordering of the quadrupoles, focus-defocus-focus in the horizontal plane and thus defocus-focus-defocus in the vertical, breaks the symmetry between the two planes after the plasma lens.
This means that the maximum vertical beam size is much larger than the maximum horizontal beam size, and that this maximum is reached inside the quadrupole with the strongest field, QDD870, as seen from Figure~\ref{fig:k-quads}.
From the thin-lens representation of a focusing quadrupole, the change in the trajectory angle of a particle is given by~\cite{wille_physics_2005}
\begin{equation}
    \Delta x' = \frac{-x}{f} = -x k L = -x k_0\left(1-\frac{\Delta p}{p_0}\right) L \,,
\end{equation}
where $x$ is the initial position of the particle, $k_0$ the nominal energy focusing strength, $L$ the quadrupole length, $p_0$ the reference forward momentum, and $\Delta p$ the particle momentum offset from the reference.
The sensitivity of the trajectory angle to the energy offset $\Delta p$ can then be found as:
\begin{equation}
    \frac{\mathrm{d}x'}{\mathrm{d}\Delta p} = 
    \frac{x k_0 L}{p_0}\,.
\end{equation}
Consequently, the higher the strength of magnet $k_\mathrm{QDD870}$, and the larger the beam size $\sigma$ (and corresponding larger particle offsets inside the magnet), the greater the sensitivity to energy errors $\Delta p$.
This then results in a larger phase space and thus larger emittance.
A more detailed discussion about how the emittance is affected by chromatic effects can be found in~\cite{lindstrom_design_2016}.

In addition to energy spread, the minimum spot size will also be limited by multiple Coulomb scattering within the target~\cite{reaz_sharp_2022}.
This effect affects both the transverse and longitudinal spot size, however, it diminishes with increasing energy.
As a result, the track densities and thus doses at the focal points, indicated in Figure~\ref{fig:trackDensity}, will be reduced.
However, the track densities before and after the focal point will remain suppressed, and consequently the beneficial dose contrast will be preserved.
The effect of scattering and the resulting dose distributions will be studied in more detail in the future, using the Geant4~\cite{agostinelli_geant4simulation_2003} software package for radiation transport Monte-Carlo simulations.

The SHARP technique focuses the beam into a small point at the target, while spreading the beam over a large area for tissues located in front of or behind the target.
In other words: The dose is highly concentrated at the target, due to the density of particle tracks there, as illustrated by Figure~\ref{fig:trackDensity}.
This would enable high-precision targeting of small tumors, and the delivery of high doses to targets near sensitive organs lateral to the beam without exposing those organs.
However, for irradiations of large volumes and targets with cross-sections similar to that of the incoming beam, the cones of incoming and outgoing beam would mostly overlap, resulting in a significant dose accumulation to surrounding tissues and a reduction in dose contrast.
This must be taken into account for dose planning.

It is natural to compare the SHARP technique with intensity modulated radiotherapy (IMRT) or intensity modulated particle therapy (IMPT)~\cite{cho_intensity-modulated_2018}, and particle arc therapy (PAT)~\cite{mein_particle_2024}.
These are some of the most advanced clinically available techniques, all with the goal of reducing the dose behind and in front of the target by approaching it from multiple angles.
However for these techniques, the resolution is limited by the movement of the patient during the time needed for the machine to mechanically move around.
This also increases the total treatment time, making it unsuitable for FLASH techniques which require sub-millisecond delivery.
These techniques also typically use relatively parallel beams, which enlarges the minimum achievable spot size.
However, these techniques may be complimentary and could potentially be combined to exploit their respective advantages.

A potential application of strongly focused beams is spatially fractionated radiotherapy (SFRT)~\cite{asperud_spatially_2021,yan_spatially_2019}.
Unlike conventional radiotherapy, SFRT delivers highly heterogenous dose distributions, characterized by localized high-dose regions (peaks) separated by low-dose regions (valleys).
This dose inhomogeneity has shown strong potential to improve normal tissue sparing while maintaining tumor control through immune responses stimulated by the high-dose peaks~\cite{bergeron_non-homogenous_2024,prezado_tumor_2019}.
This application would strongly utilize the very specific dose delivery enabled by the SHARP technique.

For the generation of large round beams, the results shown here are restricted to linear lenses and Gaussian beams.
However, for FLASH irradiation of large areas, non-Gaussian uniform beam profiles are wanted at the target in order to create uniform dose profiles~\cite{robertson_demonstration_2026,korysko_methods_2022,angoletta_deft_2025}.
We want to investigate whether a variation of our system can produce such transverse profiles by employing an APL with a non-linear focusing force to magnify the beam.
This type of field can be achieved by using a light gas such as helium in the APL~\cite{carl_andreas_lindstrom_emittance_2019,van_tilborg_nonuniform_2017,rockemann_direct_2018}.
A fully magnetic way to tailor the beam distribution would be highly beneficial, as it would preserve beam intensity and avoid stray radiation.

If a large-area beam is desired but it is not required to be parallel, the focusing quadrupoles can be omitted.
By extrapolating the expanding beam envelope curves coming out of the plasma lens in Figure~\ref{fig:APLsigma2}, it is clear that even larger beam sizes or shorter lattices can be achieved.

One of the unknowns about FLASH therapy is to which degree it is possible to combine it with pencil beam scanning, and exactly how the FLASH effect depends on the temporal structure, spot size, and scanning pattern of the pencil beams~\cite{folkerts_framework_2020}.
By changing the APL gradient and final focus quadrupole settings, pencil beams with a wide range of spot sizes and convergence angles can be produced, which could also be scanned.
This could thereby help in answering these important research questions.

Both strongly focusing and high-$\beta$ optics can also be made using quadrupole magnets alone, which is the baseline plan for the new beamline at the CLEAR facility~\cite{dyks_design_2022}.
As part of the SHARP project, we will compare the performance of a SHARP-type hybrid focusing system with that of pure quadrupole system.
Previous experience has shown that it is difficult to achieve high-quality symmetric sharp focusing in both planes when using only quadrupoles in a short lattice~\cite{kokurewicz_experimental_2021,whitmore_cern-based_2024}.

This study has mainly focused on electrons, for which improved dose conformity can be achieved using three-dimensional spot scanning based on the target geometry, especially at high beam energies~\cite{kokurewicz_focused_2019}.
However, electron beams will still leave a notable dose behind the target due to their inherent interactions with matter, compared to proton beams. 
That said, the focusing technique should also be applicable to protons, which exhibit a true Bragg peak.
State-of-the-art proton therapy leverages this property, combined with spot scanning, to achieve high dose conformity.
Applying the proposed technique to protons allows for a reduction in spot size near the Bragg peak, effectively further improving dose conformity, while simultaneously the beam remains more diffuse where it enters the patient.
A further reduction in small spot size can be reached with high energy protons, reducing the effect of scattering, and setting the spot depth by beam geometry and reducing the dependence of tissue density~\cite{reaz_sharp_2022}.
However, for such spots the protons will behave similarly to electrons, loosing the benefits of the Bragg peak but also becoming less sensitive to density inhomogeneities~\cite{lagzda_influence_2020}.
This flexibility might provide ``the best of both worlds'' within a single facility.
We plan to study the application to protons as part of the SHARP project.
One important technical issue to address is how to adapt the APL technology, which currently operates with pulses where the current is changing on the sub-microsecond timescales, to the quasi-continuous proton therapy beams typically generated by cyclotrons.

One potential challenge for using an APL in a final focus system, is the additional ``passive'' plasma wakefield focusing induced by the bunch when traveling through the lens~\cite{su_plasma_1990}.
This can perturb --- and potentially overwhelm --- the focusing provided by the lens current $I_\mathrm{APL}$~\cite{lindstrom_analytic_2018}, degrading the beam emittance~\cite{kim_witness_2021} through the addition of higher-order nonlinear field terms.
This could affect the effective (de-)focusing strength of the lens which is normally set by the lens current, altering the resulting focal spot size and convergence angle.
However, we see from Figure~\ref{fig:k-quads} that the focus point is determined by the quadrupole strengths, and does not have a strong variation with $I_\mathrm{APL}$.
Furthermore, the plasma lens is used in defocusing mode, which avoids the rapid increase in beam density around the APL exit that occurs for converging beams.
In~\cite{kim_witness_2021} this density spike was shown to be problematic, because it more readily excites wakefields in the plasma density downramp around the APL exit.

While the SHARP project is primarily targeting applications in radiotherapy, the technique may be generalised to final focusing systems for particle colliders~\cite{white_beam_2022}.
For potential future linear colliders such as CLIC~\cite{clic_collaboration_multi-tev_2012}, ILC~\cite{behnke_international_2013}, or HALHF~\cite{foster_proceedings_2025}, the final focus is a very large and complex system, making up a significant fraction of the overall machine.
The authors are therefore interested in seeing whether the use of a diverging plasma lens could be helpful in making these final focus systems simpler, making it more likely that such machines will be constructed, for the benefit of particle physics research and pushing the frontiers of particle accelerators.

\section{Conclusions}

The SHARP project is investigating how a compact hybrid focusing system for medical applications of particle accelerator beams can be made using a combination of active plasma lenses and quadrupole magnets.
In this paper, the basic charged particle beam optics concepts are presented.
It was shown that such a compact system can reach small beam sizes and steep convergence angles with high precision, with meter-scale throw distances between the exit of the final magnet and the target.
Furthermore, the optics is flexible with a movable focal point.
The design is robust against energy spread, with emittance increasing slowly for low plasma lens currents, while performance gains in beam size and convergence happen immediately.

It has also been shown that this system can generate large, parallel beams for broad beam irradiation.
Since these beams are formed magnetically, they provide a much cleaner radiation field with higher intensity than beams enlarged by passive scattering.
This is particularly advantageous for FLASH therapy.

We are working on refining the concept further, through simulations exploring simplified magnet lattices with e.g.\ only two final quadrupoles, sensitivity to non-ideal initial beam parameters, and transverse spot scanning using steering magnets.
Also under investigation are the effect of scattering when passing through beam-pipe vacuum windows, air, and a water phantom surrounding the target, as well as effects from unwanted plasma wake field effects in the APL.
We are currently conducting experimental tests using the CLEAR electron beam, in order to validate the concept.

Through this, we are in the near term looking to create a valuable tool for research into SFRT and FLASH, and eventually enable higher precision radiotherapy.
As a bonus, the hybrid focusing technique may also be valuable for particle collider applications.

\section*{Acknowledgments}
This work was supported by the Research Council of Norway, grant No. 353317.


\printbibliography

\end{document}